\documentclass[preprint,12pt]{elsarticle}

\usepackage{amssymb}
\usepackage{amsmath}
\usepackage{graphicx}
\usepackage{subfig}
\usepackage{float}
\usepackage{xcolor}
\usepackage[colorlinks=true,
            urlcolor=black,
            linkcolor=black,
            filecolor=black,
            anchorcolor=black,
            citecolor=black
]{hyperref}

\begin{document}

\begin{frontmatter}



  \title{Scaling of average trapping time and average weighted shortest path on a residual multi-weighted crystal network}


  \author[author1]{Bingyan Xie}
  \affiliation[author1]{organization={School of Applied Mathematics}, 
    addressline={Nanjing University of Finance and Economics},
    city={Nanjing},
    postcode={210023},
    state={Jiangsu},
    country={CHN}}
  \author{Bo Wu \corref{cor1}}  
  \ead{bowu8800@nufe.edu.cn}
  \cortext[cor1]{Corresponding author}

  \begin{abstract}
    This article constructs the residual network after some regions were damaged and detached from the original crystal network. This residual crystal network simulates the situation where parts of a computer system or power system failed after been attacked in real life. Furthermore, we assign multiple weight factors to the edges in the network, exhibiting mixed weight growth. By using the symmetry and self-similarity of the network structure, the analytical expression for the average trapping time and the average weighted shortest path on this network are solved, and the numerical results are given by taking the residual hexagonal crystal network as an example. By analyzing the network and studying the topological properties, we show the robustness of the network structure and find the residual network is more efficient for communication between nodes.
  \end{abstract}




\end{frontmatter}


\section{Introduction}
\label{sec:intro}

In recent years, complex network has become an important research hotspot, which refers to a type of network with properties such as self-organization, small-world, scale-free and so on \cite{watts1998collective, barabasi1999mean, milo2002network}. The theoretical achievements and applicable value of complex network have attracted widespread attention of scholars from various fields. This is because complex network has close connection with real world system, like traffic system, neural system, signal transmission system and so on \cite{albert2000error, boccaletti2006complex, doi:10.1137/S003614450342480}. Complex network can precisely describe many behaviors by using nodes of network to represent different elements of system, and links between nodes to represent the interactions between elements. The rise of complex network has opened up new ideas and perspectives for scientific research in various disciplines. For example, a new method has been proposed to determine the optimal location of microgrids in the power system \cite{4272326}.

Through the study of complex network, Song et al have unveiled the presence of self-similarity \cite{song2005self}. Many scholars have delved into self-similar network such as SG network and Koch network, and further investigating their topological properties including clustering coefficients, degree distributions and so on, which bear great significance to the analysis of dynamic characteristics \cite{wang2019scale} \cite{zhang2021mean}.

The trapping problem has become a hot topic among them. The indicator of this problem is average trapping time(ATT), which means the average time for a node to be captured by the trap through random walk, reflecting the absorption efficiency of the trap and information transmission efficiency of the entire network. Scholars have investigated the trapping problem on self-similar network such as hierarchical network, (u,v)-flowers, pseudofractal network, and they have obtained the analytical expression. A great number of researches showed that ATT usually increases linearly with the network size, but sometimes it increases sub-linearly due to the network structure and weight factor. Moreover, changing the random walk rule on the network generally does not affect the main exponent of the analytical expression, but it affects the coefficients \cite{sun2016scaling, xing2019trapping, wu2020average} .

The shortest path problem is another important issue. The indicator of this problem is average weighted shortest distance(AWSP), which means the average distance between a pair of two nodes in the network, describing the small-world property of self-similar network. It is commonly used to calculate the network diameter and reflect the evolutionary trend of the entire network. There are many studies focus on it within self-similar network such as Cayley network, Vicsek network. Many researches showed that AWSP usually increases proportionally with the number of iteration. However, when the weight is assigned to the network, AWSP usually increases with the growth of weight factor. The number of iteration may no longer has a significant impact on it, and it tends to a constant value \cite{niu2018scaling} \cite{lei2021node}.

In addition, some scholars have investigated above problems in the weighted network which is more relative to the real world, because the status of elements in system are sometimes different and the tightness of connections vary greatly. Based on a class of weighted-crystal network proposed by Li \cite{li2021scaling}, we create the remaining network after parts of its area were destroyed and detached from the whole. This residual network could simulate many realistic situations where the integrality of the network is compromised. We analyze ATT and AWSP re the network was damaged, which could show how the change of structure influences the network dynamics. In addition, we assign multiple weight factors to the edges in the network, exhibiting mixed weight growth with iteration. Comparing with single weight factor in the original network, the variation range of weight is extended. By using the symmetry and self-similarity of the network structure, the iterative relationship is obtained so that the calculation can be simplified. The analytical expression for the ATT and AWSP are solved, and the specific numerical results are given by using the residual hexagonal crystal network as an example.

The chapter arrangement of this article is as follows: Section \ref{section2} introduces the structure and iteration method of the network. Section \ref{section3} presents the analytical expression of ATT and gives numerical results by using residual hexagonal crystal as an example. Section \ref{section4} presents the analytical expression of AWSP and gives numerical results by residual hexagonal crystal. In Section \ref{section5}, schematic diagrams are demonstrated to reveal the trend of these two topological properties.

\section{The structure of residual multi-weighted crystal network}
\label{section2}

\begin{figure}[H]
  \centering
  \subfloat{
    \includegraphics[width=0.2\textwidth]{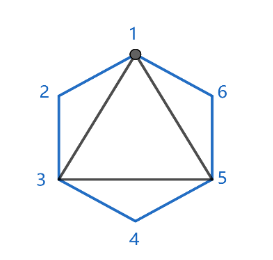}
  }
  \subfloat{
    \includegraphics[width=0.25\textwidth]{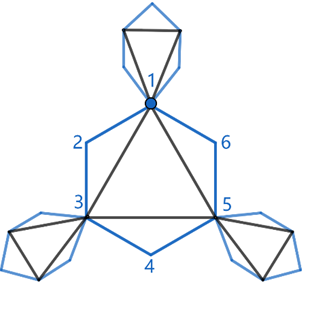}
  }
  \subfloat{
    \includegraphics[width=0.45\textwidth]{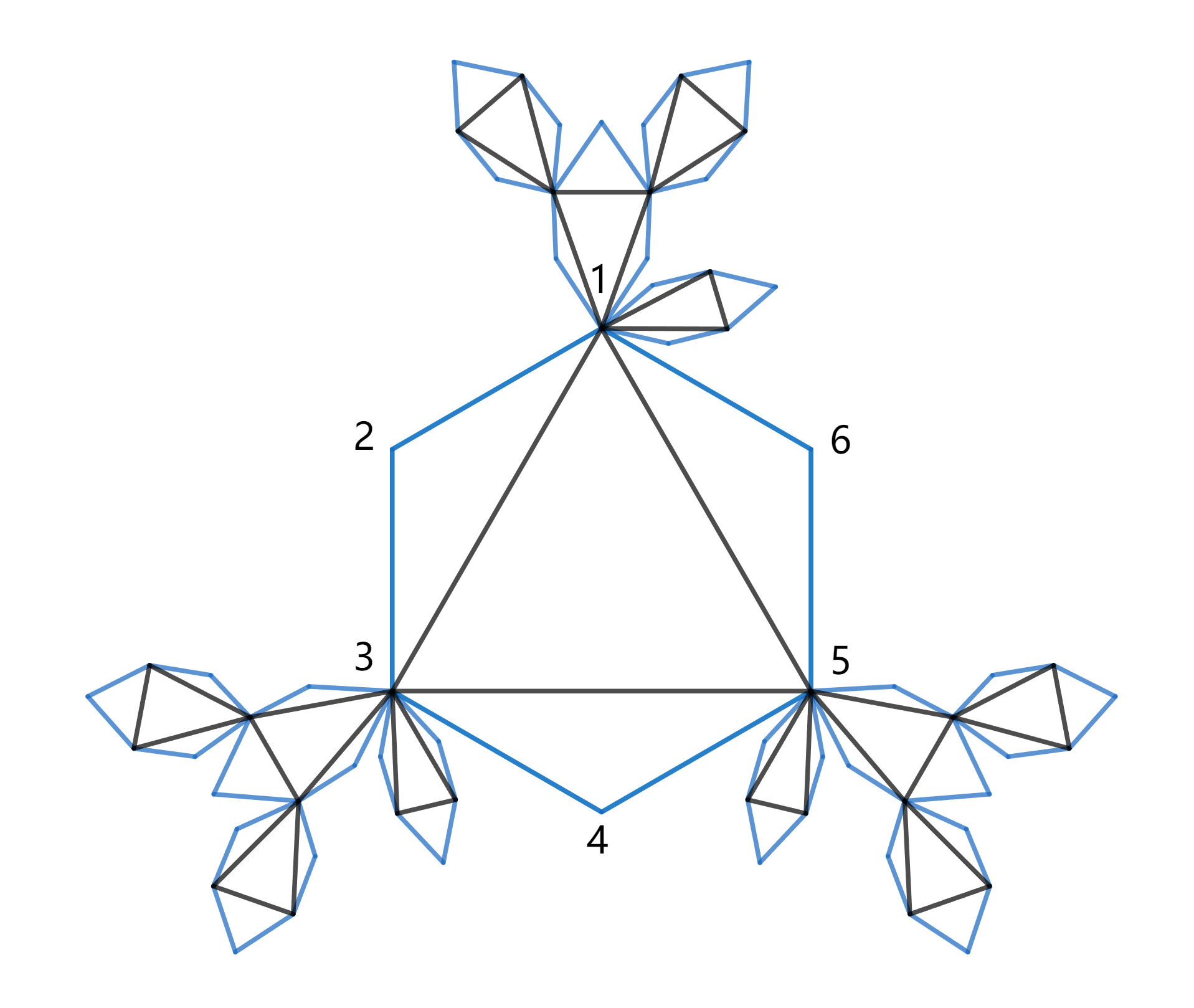}
  }
  \caption{\centering{Iterative Diagram of the original crystal network(n=3).}}
  \label{figure1}
\end{figure}

The original network has an even number of nodes(2n), all nodes in the initial graph are regarded as hub nodes, but only odd nodes have the ability to iterate(see reference \cite{li2021scaling}). The iterative rule of the network is as follows: when $t\ge 1$, $G_t$ is obtained by connecting node 1 in $G_{t-1}$ to all odd nodes in $G_0$ as their respective suffix domains. As shown in Figure \ref{figure1}, only odd nodes have their appendage region (when $t\ge 1$). By this definition, name odd nodes as primary nodes and even nodes as secondary nodes. An iteration diagram of crystal network is given (taking n = 3 as an example).

Based on this original crystal network, we are aimed to create a residual network which can describe the attacked situation. In this article, we stimulate that node 1 has been attacked at the beginning, meaning that node 1 lost its iterative capability and the appendage region of it was detached from the whole network. Furthermore, we are intend to create the network which is more relative to the real system, so the multiple weight factors are considered.

The iterative and weighted process is as follows:
\begin{enumerate}[(a)]
  \item In $G_t$, name the central region $A_0^t$, and the n-1 suffix domains $A_i^t$($i = 1,2,3\cdots {n-1}$). We consider multiple weight factors $\theta _i$ ($i=1,2,3\cdots {n-1}$), each weight factor is assigned to $A_i^t$ in sequence.
  \item Every $A_i^t$ is obtained by duplicating $G_{t-1}$, and the edge weight is scaled by $\theta_i$ respectively, so $G_{t-1}$ is copied {n-1} times from {t-1} to t generation. We name the edge weight in $A_i^t$ to be  $A_i^t$ and the edge weight in $G_{t-1}$ to be $\omega G_{t-1}$, then we have $\omega _{A_i^t} = \theta _i\cdot \omega _{G_{t-1}}$.
  \item Then node 1 of each duplication is combined with hub node which has the iterative ability in $G_0$ (node $3,5,\cdots {2n-1}$).
\end{enumerate}

\begin{figure}[H]
  \centering
  \captionsetup[subfloat]{labelsep=none,format=plain,labelformat=empty} 
  \subfloat[t = 0]{
    \includegraphics[width=0.3\textwidth]{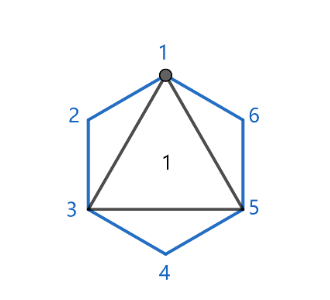}
  }
  \subfloat[t = 1]{
    \includegraphics[width=0.5\textwidth]{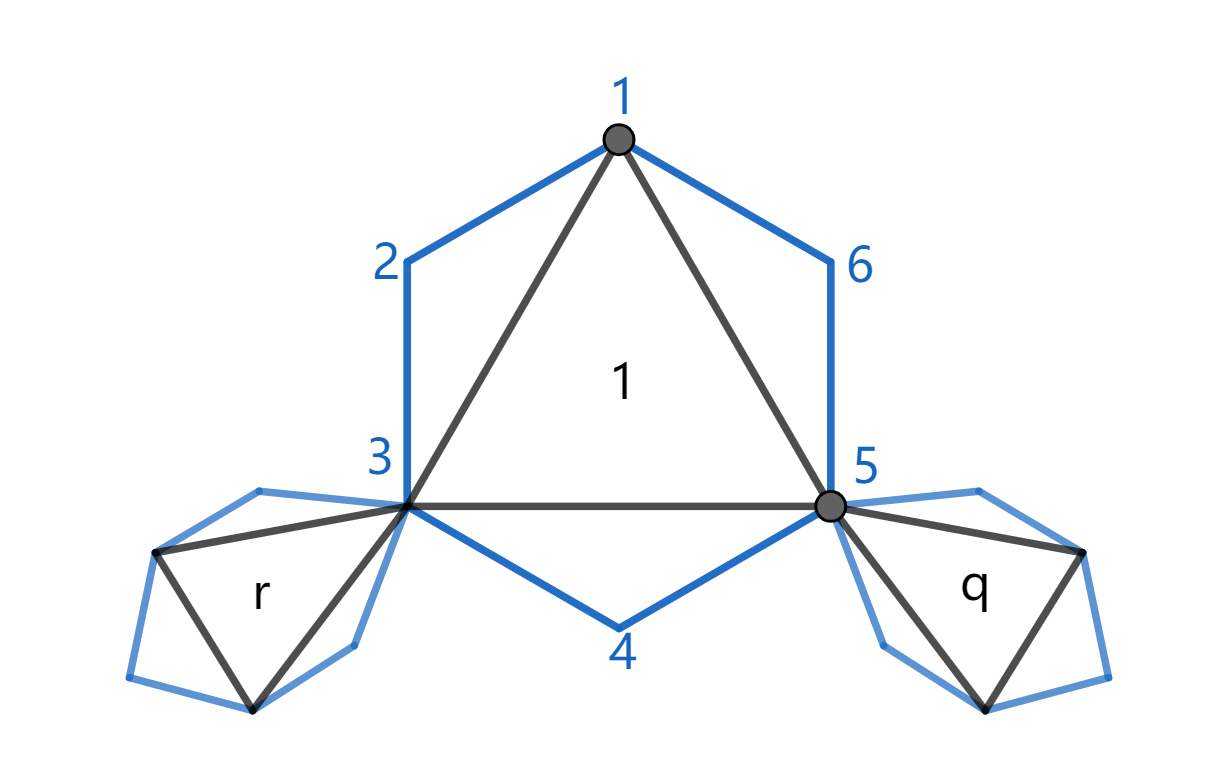}
  }
  \\
  \subfloat[t = 2]{
    \includegraphics[width=0.5\textwidth]{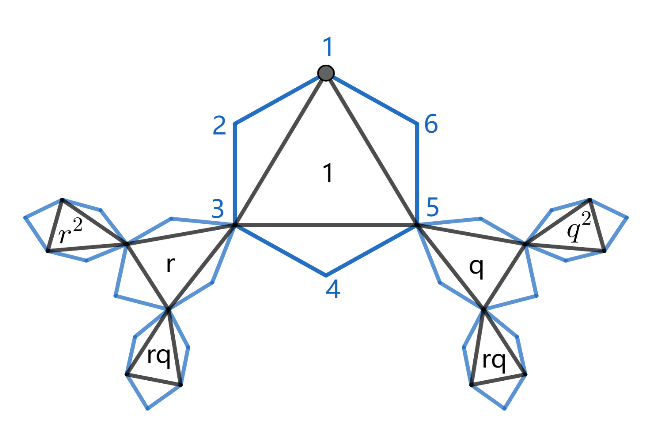}
  }
  \\
  \subfloat[t = 3]{
    \includegraphics[width=0.7\textwidth]{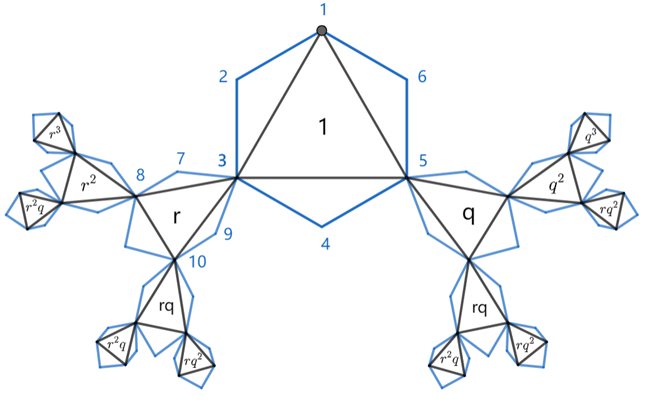}
  }
  \caption{\centering{Iterative diagram of the residual network take {n = 3} as example ($\theta _1 = r$ and $\theta _2 = q$).}}
  \label{figure2}
\end{figure}

\section{Average trapping time in residual multi-weighted crystal network}
\label{section3}
\subsection{The derivation process of average trapping time}
This section gives the derivation process of ATT. We consider a weight-based random walk in this residual multi-weighted crystal network. The random walker starts from any node in the network except the position where trap was set (in this article we set node 1 to be the trap). When a random walker reaches the trap for the first time, it is immediately captured and no longer walks on the network. Otherwise, it will continue the walking process. Each time the walker moves from one position to another depends totally on the transition probability. The transition probability from node i to node j is

\[ P_{ij}=\frac{\omega _{ij} }{s_{i} }  =\frac{\omega _{ij} }{\sum_{j\in \nu (i)}\omega _{ij} }.\]

where $\omega _{ij}$ is the weight of edge connecting node i and node j, $\nu_i$ is the node set consists of nodes directly connected to node i, $s_i$ is the strength of node i calculated by summing edge weight of nodes from $\nu_i$.

\begin{align}
  &F_{i} (t) =1 + \sum_{j\in \nu _{i} }p_{ij} F_{j} (t),\nonumber\\
  &T_{tot} (t) =\sum_{i=2 }^{N_{t} } F_{i} (t),\nonumber\\
  &\left \langle T \right \rangle _{t} =\frac{1}{N_{t}-1 } T_{tot}(t),\nonumber
\end{align}

where $F_i (t)$ means the time consumed for being captured by the trap when the walker starts from node i in $G_t$, $T_{tot} (t)$ means the time consumed for the walker traversing all possible starting position in the network $G_t$, $\left \langle T \right \rangle _{t}$ means ATT in $G_t$.

According to the network structure, it is obtained that if the walker starts from the node in suffix regions, it must pass through its own primary node odd hub node to reach the trap. What's more, according to the self-similarity of crystal network, every suffix region in $G_t$ is similar to $G_{t-1}$, and being attacked doesn't change this property. Thus, we can obtain

\begin{equation}
  T_{tot} (t)=(n-1)T_{tot} (t-1)+\frac{N_{t} -(n+1)}{n-1} \cdot \sum_{m=1}^{n-1} F_{2m+1} (t)+\sum_{m=1}^{n} F_{2m} (t),
  \label{equation1}
\end{equation}

where $F_{2m+1}(2n,t) $ means the time consumed for being captured when the walker starts from odd hub node 2m+1 ($1\le m\le n-1$) in $G_t$, $F_{2m}(2n,t) $ means the time consumed for being captured when starts from even hub node 2m ($1\le m\le n$) in $G_t$.

From the structural features of the network, it can be inferred that 

\begin{equation}
  F_{2m} (2n,t) = \left\{
  \begin{array}{ll}
    F_{2} &= 1+F_{3}(t) \\ 
    F_{2n} &= 1+\frac{1}{2} F_{2n-1}(t) \\ 
    F_{2m} &= 1+\frac{1}{2} F_{2m-1}(t)+\frac{1}{2} F_{2m+1}(t) 
  \end{array} \right.,\nonumber
\end{equation}

\begin{align}
  F_{2m+1} (2n,t) =& 1+\sum_{j\in \nu _{2m+1} }p_{2m+1,j} F_{j}(t) \nonumber\\
  =& 1+\frac{1}{4+4\theta _{i} }\left [ F_{2m-1}(t)+F_{2m}(t)+F_{2m+2}(t)+F_{2m+3}(t) \right ] \nonumber\\ &+\frac{\theta _{i} }{4+4\theta _{i} }\left [ 4F_{2m+1}(t)+F_{2}(t-1)+F_{3}(t-1)+F_{2n}(t-1)+F_{2n-1}(t-1)  \right ] \nonumber\\
  =& 2+2\theta _{i}+\frac{1}{2}\left [ F_{2m-1}(t)+F_{2m+3}(t) \right ]+\frac{\theta _{i} }{2}\left [ F_{3}(t-1)+F_{2n-1}(t-1) \right ]. \nonumber
\end{align}

To calculate $\sum_{m=1}^{n}F_{2m}(t) $, we can get

\begin{equation}
  \sum_{m=1}^{n}F_{2m}(2n,t) = n+\sum_{m=1}^{n-1}F_{2m+1}(2n+t). 
  \label{equation2}
\end{equation}

To calculate $\sum_{m=1}^{n-1}F_{2m+1}(t) $, we can get

\begin{align}
  \sum_{m=1}^{n-1}F_{2m+1}(2n,t) =& 2(n-1)+2\sum_{i=1}^{n-1}\theta _i+\left \{ \sum_{m=1}^{n-1}F_{2m+1}(2n,t)-\frac{1}{2}\left [ F_{3}(t)+F_{2n-1}(t) \right ] \right \} \nonumber\\
  &+\sum_{i=1}^{n-1}\frac{\theta _{i} }{2}\left [ F_{3}(t-1)+F_{2n-1}(t-1) \right ].
  \label{equation3}
\end{align}

Substitute Eq.(\ref{equation2}) and Eq.(\ref{equation3}) in Eq.(\ref{equation1}) to get

\begin{equation}
  T_{tot}(t)=(n-1)T_{tot}(t-1)+\frac{N_{t}-2 }{n-1}\sum_{m=1}^{n-1}F_{2m+1}(2n,t)+n.
  \label{equation4} 
\end{equation}

From Eq.(\ref{equation4}), we need to solve $\sum_{m=1}^{n-1}F_{2m+1}(t) $ in order to solve ATT. According to the walking process, the trapping time for odd hub node {2m+1} can be transformed to the sum of trapping time for the left neighbored odd hub node 2m-1 and the right neighbored odd hub node 2m+3. By recursive relationship, $\sum_{m=1}^{n-1}F_{2m+1}(t) $ can be transformed to the sum of trapping time for the first left neighbored node 3 and the last right neighbored node {2n-1}. That is, $F_3 (t)+F_{2n-1} (t)$ need to be solved.

We simplify Eq.(\ref{equation4}) to get that
\begin{equation}
  F_{3}(t)+F_{2n-1}(t)=4(n-1)+4\sum_{i=1}^{n-1}\theta _{i }+\sum_{i=1}^{n-1}\theta _{i}\left [ F_{3}(t-1)+F_{2n-1}(t-1) \right ]. \nonumber
\end{equation}

Let define M(t) to be
\begin{equation}
  M(t)=F_{3}(t)+F_{2n-1}(t).
  \label{equation5}
\end{equation}

Then we have the analytical expression of M(t)
\begin{align}
  M(t) &= 4(n-1)+4\sum_{i=1}^{n-1}\theta _{i}+\sum_{i=1}^{n-1}\theta _{i}\cdot M(t-1) \nonumber\\ 
  &=4(\sum \theta _{i} )^{t}(n-1)+4\frac{(\sum \theta _{i} )^{t}-1 }{(\sum \theta _{i})-1 } \left [ n-1+(\sum \theta _{i} ) \right ]. \nonumber
\end{align}

Now we have proved ATT can be solved by recursive way. 
\subsection{The calculation of ATT when {\textbf{n} = 3}}
Let take the residual hexagonal crystal network (n = 3) as an example to give the specific calculation process of ATT. In this case, we stimulate the situation where the original network (which has the similar structure with a benzene ring molecule) has been attacked and appendage region of node 1 was invalid. When n = 3, name the two weight factors in $G_t$ are $\theta _1=r$ and $\theta _2=q$.

The number of nodes in $G_t$ is
\[ N_{t}=5\cdot 2^{t+1}-4. \]

From Eq.(\ref{equation1}), the time consumed for the walker traversing all possible starting positions in the network is
\[ T_{tot}(t)=2T_{tot}(t-1)+\frac{N_{t}-4}{2}\left [ F_{3}(t)+F_{5}(t) \right ]+F_{2}(t)+F_{4}(t)+F_{6}(t). \]

From the structural features of the network, it can be inferred

\begin{equation}
  \left\{
  \begin{array}{ll}
    F_{2}(t) = &1+\frac{1}{2}F_{3}(t) \\  
    F_{4}(t) = &1+\frac{1}{2}F_{3}(t)+\frac{1}{2}F_{5}(t) \\
    F_{6}(t) = &1+\frac{1}{2}F_{5}(t)
  \end{array} \right..\nonumber
\end{equation}

Then we have
\[ T_{tot}(t)=2T_{tot}(t-1)+(5\cdot 2^{t}-3)\left [ F_{3}(t)+F_{5}(t)\right ]+3. \]

By the definition in Eq.(\ref{equation5}), when n = 3, $M(t)=F_3 (t)+F_5 (t)$.Then $T_{tot} (t)$ can be expressed to
\begin{align}
  T_{tot}(t) =& 2T_{tot}(t-1)+(5\cdot 2^{t}-3)\cdot M(t)+3 \nonumber\\
=& 19\cdot 2^{t}+3\cdot (2^{t}-1)+5\cdot 2^{t}\left [ M(1)+\cdots +M(t)\right ] \nonumber\\ 
&-3\cdot \left [ 2^{t-1}M(1)+\cdots +M(t)\right ].
\label{equation6}
\end{align}

As shown in Figure \ref{figure2}, it can be inferred
\begin{equation}
  \left\{
  \begin{array}{cl}
    F_{7}(t) &= F_{2}(t-1)+F_{3}(t) \\
    F_{8}(t) &= F_{3}(t-1)+F_{3}(t) \\  
    F_{9}(t) &= F_{5}(t-1)+F_{3}(t) \\
    F_{10}(t) &= F_{6}(t-1)+F_{3}(t)
  \end{array} \right..\nonumber
\end{equation}

By the rule of weight-based random walk
\begin{align}
  F_{3}(t) &= 1+\frac{1}{4+4r}\left ( F_{2}(t)+F_{4}(t)+F_{5}(t)\right )+\frac{r}{4+4r}\left ( F_{7}(t)+F_{8}(t)+F_{9}(t)+F_{10}(t)\right ) \nonumber\\
&= 2+2r+\frac{1}{2}F_{5}(t)+\frac{r}{2}\left [ F_{3}(t-1)+F_{5}(t-1) \right ]. \nonumber
\end{align}

Similarly, we can obtain
\[ F_{5}(t)=2+2q+\frac{1}{2}F_{3}(t)+\frac{q}{2}\left [ F_{3}(t-1)+F_{5}(t-1)\right ]. \]

Sum up $F_{3}(t)$ and $F_{5}(t)$ to get M(t)
\[ M(t)=(r+q)^{t}\cdot \frac{12(r+q)}{(r+q)-1}-\frac{4\left [2+(r+q)\right ] }{(r+q)-1}, \]
if $(r+q)\ne 1,2$.

Substitute the above results in Eq.(\ref{equation6}) and simplify it, we have three cases of the result

\begin{enumerate}[(i)]
  \item If $(r+q)\in (0,2)$ and $(r+q)\ne 1,2$
  \begin{align}
    T_{tot} =& 11\cdot 2^{t+1}-3 -\frac{12\cdot \left [ 2+(r+q)\right ] }{(r+q)-1} \nonumber\\ 
    &+ 5\cdot 2^{t+1}\left [ \frac{\frac{6}{5}\cdot \left [ 2+(r+q)\right ] }{(r+q)-1}+\frac{6\cdot (r+q)}{(r+q)-1}\cdot \frac{(r+q)^{t+1}-(r+q)}{(r+q)-1}-\frac{2\cdot \left [ 2+(r+q)\right ] }{(r+q)-1}\right ] \nonumber\\ 
    &- 2^{t}\cdot \frac{36\cdot (r+q)}{(r+q)-1}\cdot \frac{\frac{r+q}{2}-(\frac{r+q}{2})^{t+1}}{1-\frac{r+q}{2}}. \nonumber
  \end{align}

  \item If (r+q)=1
  \[ T_{tot} = 30\cdot t^{2}\cdot 2^{t}+70\cdot 2^{t}\cdot t-38\cdot 2^{t}+21. \]

  \item If (r+q)=2
  \[ T_{tot} = 15\cdot 2^{2t+4}-152\cdot 2^{t}\cdot t-170\cdot 2^{t}-51. \]

\end{enumerate}

Take the limit value of the result in the above equation, i.e. $t\to \infty $

\begin{equation}
  \left \langle T \right \rangle_{t} \sim \left\{
  \begin{array}{ll}
    \frac{2\cdot \left [ 2+(r+q)\right ]}{1-(r+q)}\cdot t  &,(r+q)\in (0,1)  \\
    \frac{2\cdot \left [ 2+(r+q)\right ]}{1-(r+q)}\cdot t+6\cdot (r+q)^{t}  &,(r+q)\in (1,2)  \\  
    3\cdot t^{2}+7t  &,(r+q)=1  \\
    3\cdot 2^{t+3}-\frac{3}{5}t   &,(r+q)=2
  \end{array} \right.\nonumber
\end{equation}

\section{Average weighted shortest path in residual multi-weighted crystal network}
\label{section4}
\subsection{The derivation process of AWSP}
This section gives the derivation process of AWSP. AWSP means the average distance between two nodes from any possible combination. Let's define $d_{ij} (t)$ to be the distance between node i and node j in $G_{t}$, $D_{tot} (t)$ to be the total distance traversing all possible nodes combination, $\lambda _{t}$ to be AWSP in $G_{t}$.

\[ D_{tot}(t)=\sum_{\substack{i\ne j \\ i,j\in G_{t}}}d_{ij}(t), \]
\[ \lambda _{t}=\frac{2}{N_{t}\cdot (N_{t}-1)}\cdot D_{tot}(t). \]

There is three cases for possible combination of two nodes:
\begin{enumerate}[(i)]
  \item both nodes are from the same suffix region $A_{i}^t$;
  \item both nodes are from the central region $A_{0}^t$;
  \item two nodes are from the different regions. What's more, according to the iteration of the network, every edge in $G_{t}$ is scaled by weight factors(when $t\ge 1$), so the distance is also scaled in this way. We can sum the distance of all nodes combinations according to the regions where two nodes belong
  \begin{equation}
    D_{tot}(t)=\sum_{i=1}^{n-1}\theta _{i}D_{tot}(t-1)+D_{tot}(0)+\pi _{t},
    \label{equation7}
  \end{equation}
  where $ D_{tot}(0)=\frac{n}{2}(n^{2}+2n-1). $
\end{enumerate}

In the above equation, three items represent the sum of distance to three cases. The third item $\Omega _t$ which means two nodes are from the different regions can be classified into two different situations: let $\Omega _t^{A_mA_0}$ to represent that one node is from the suffix regions while the other is from the central region, let $\Omega_t^{A_mA_n}$ to represent that two nodes are from the different suffix regions.
\begin{equation}
  \Omega _{t}=\sum_{m=1}^{n-1}\Omega _{t}^{A_mA_0}+\sum_{m=1}^{n-2}\sum_{p=1}^{n-m-1}\Omega _{t}^{A_mA_{m+p}}.
  \label{equation8}
\end{equation}

To simplify the calculation, let define $\Delta _{t}$ to be an intermediate variable representing the distance between all nodes and node 1 in $G_{t}$
\begin{equation}
  \Delta _{t}=\sum_{\substack{i\in G_{t}\\ i\ne 1}}d_{i1},
  \label{equation9}
\end{equation}
where $\Delta _0=\frac{n(n+1)}{2}$.

Then we have
\begin{align}
  \Omega _{t}^{A_mA_0} &= \sum_{i\in A_m}\sum_{j\in A_0}d_{ij}=(N_{0}-1)\cdot \theta _{i}\cdot \Delta_{t-1}+(N_{t-1}-1)\cdot \Delta _0, \\
  \sum_{m=1}^{n-1}\Omega _{t}^{A_mA_0} &= (N_{0}-1)\cdot \sum\theta _{i}\cdot \Delta_{t-1}+(n-1)\cdot (N_{t-1}-1)\cdot \Delta _0, \nonumber \\
  \Omega _{t}^{A_mA_{m+p}} &= \sum_{i\in A_m}\sum_{j\in A_{m+p}}d_{ij}=(N_{t-1}-1)\cdot (\theta _{i}+\theta _{j})\cdot \Delta_{t-1}+p\cdot (N_{t-1}-1)^2,
\end{align}

\begin{equation}
  \sum_{m=1}^{n-2}\sum_{p=1}^{n-m-1}\Omega _{t}^{A_mA_{m+p}}=\left\{
  \begin{array}{lc}
  (N_{t-1}-1)\cdot 2\sum \theta _i\cdot \Delta _{t-1}+(N_{t-1}-1)^2 \cdot 2(1(n-2) \\+2(n-3)+\cdots +\frac{n-1}{2}\cdot \frac{n}{2}) &, \text{if n is even} \\
  (N_{t-1}-1)\cdot 2\sum \theta _i\cdot \Delta _{t-1}+(N_{t-1}-1)^2\cdot [2(1(n-2) \\+2(n-3)+\cdots +\frac{n+1}{2}\cdot \frac{n-3}{2})+(\frac{n-1}{2} )^2] &, \text{if n is odd}
  \end{array} \right.\nonumber
\end{equation}

\subsection{The calculation of AWSP when {\textbf{n} = 3}}
Let take the residual hexagonal crystal network (n = 3) as an example to give the specific calculation process of AWSP.

When n = 3
\begin{align}
  \lambda _{t} &= \frac{1}{25\cdot 2^{2t+1}-5\cdot 2^{t+2}-25\cdot 2^{t}}\cdot D_{tot}(t), \nonumber \\
  D_{tot}(t) &= rD_{tot}(t-1)+qD_{tot}(t-1)+D_{tot}(0)+\Omega _{t}, \nonumber \\
  \Omega _{t} &= \Omega _{t}^{A_1A_0}+\Omega _{t}^{A_2A_0}+\Omega _{t}^{A_1A_2} \nonumber \\ 
  &= (r+q)\cdot 5\cdot 2^t\cdot \Delta _{t-1}+25\cdot 2^{2t}+5\cdot 2^{t+1}-35. \nonumber
\end{align}

Simplify to have
\[ D_{tot}(t)=(r+q)\cdot D_{tot}(t-1)+(r+q)\cdot 5\cdot 2^t\cdot \Delta _{t-1} +25\cdot 2^{2t}+5\cdot 2^{t+1}-14, \]
where $D_{tot}(1)=81\cdot (r+q)+106$, $D_{tot}(0)=21$.

Now the calculation of $D_{tot} (t)$ can be transform to the solution of $\Delta _{t}$. According to the definition of $\Delta _{t}$ in Eq.(\ref{equation9})
\begin{align}
  \Delta _{t} &= \sum_{i\in G_{t}}d_{i1}=\sum_{i\in A_{1}}d_{i1}+\sum_{i\in A_{0}}d_{i1}+\sum_{i\in A_{2}}d_{i1} \nonumber \\
  &= (r+q)\Delta _{t-1}+2(N_{t-1}-1)+\Delta _0 \nonumber \\
  &= (r+q)\Delta _{t-1}+10\cdot 2^t-4 \nonumber
\end{align}
where $\Delta _0 = 6$.

Then we can get the iterative expression of $\Delta _t$
\begin{equation}
  \Delta _t = \left\{
  \begin{array}{ll}
    (6-\frac{4}{(r+q)-1}-\frac{20}{2-(r+q)})\cdot (r+q)^t+\frac{20}{2-(r+q)}\cdot 2^t+\frac{4}{(r+q)-1} &, (r+q)\in (0,2),(r+q)\ne 1 \\
    5\cdot 2^{t+2}-4t-14 &, (r+q) = 1 \\
    (5t + 1)\cdot 2^{t+1}+4 &, (r+q) = 2
  \end{array} \right.\nonumber
\end{equation}

Substitute the results we get above into the expression of $\lambda _t$ and take the limit value, i.e. $t \to \infty $

\begin{equation}
  \lambda _{t} \sim \left\{
  \begin{array}{ll}
    \frac{2+3\cdot \theta }{4-2\cdot \theta } &,(r+q)\in (0,2),(r+q)\ne 1 \\
    2 &, (r+q) = 1 \\
    t &, (r+q) = 2
  \end{array} \right.\nonumber
\end{equation}

\section{Conclusions}
\label{section5}
\begin{figure}[H]
  \centering
  \includegraphics{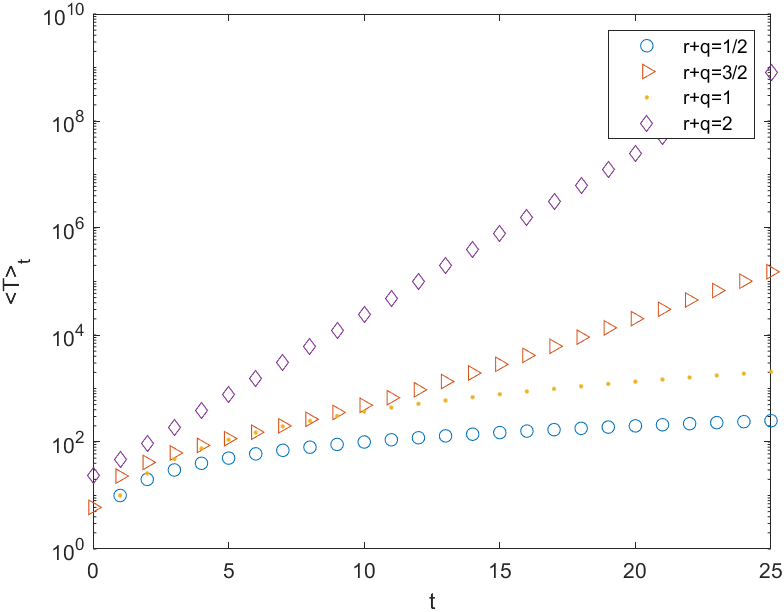}
  \caption{\centering{The asymptotic behavior of $\left \langle T\right \rangle _t$ with t on a semilogarithmic scale for r+q = $\frac{1}{2}$,1,$\frac{3}{2}$,2.}}
\end{figure}

\begin{figure}[H]
  \centering
  \includegraphics{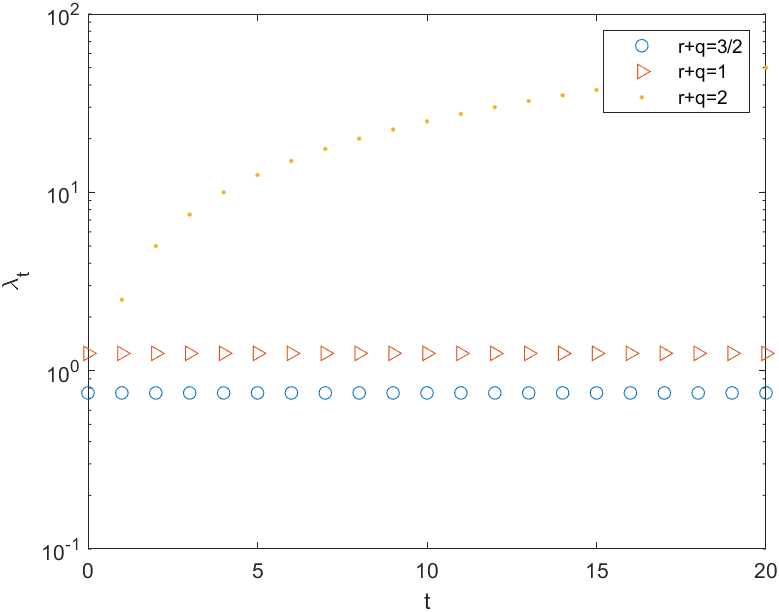}
  \caption{\centering{The asymptotic behavior of $\lambda _t$ with t on a semilogarithmic scale for r+q = 1,$\frac{3}{2}$,2.}}
\end{figure}

In conclusion, we construct the residual crystal network with the appendage domain connected to node 1 was detached from the original network. Furthermore, we assign multiple weight factors to the edges in the network. In this case, the attacked situation is analyzed .The numerical result of ATT reflects that except for the case where r+q=2, ATT shows a sublinear growth trend on a semilogarithmic scale. Weight assignment decreases ATT by lifting the probability of choosing the central area path(also the 'right path') through the walking process. By analyzing the unweighted case, we find that transformation efficiency is higher in this residual crystal network compared with the original crystal network in reference \cite{li2021scaling}.The numerical result of AWSP reflects that when the network is weighted, AWSP increases with the growth to the sum of weight factors but no longer increases with iteration. When the weight factors are determined, it tends to be constants. When the network is unweighted, AWSP increases proportionally with the iteration time t. By analyzing the unweighted case, we find that the average distance between nodes is smaller in this residual crystal network compared with the original crystal network in reference \cite{li2021scaling}. What's more, we analyze the structure of the residual crystal network and find that it still has the self-similarity, which indicates the robustness of the original crystal network. The numerical results of two topological properties indicate that this residual network is more efficient than the original network.





\newpage

\bibliographystyle{elsarticle-num-names}
\bibliography{reference}
\biboptions{numbers,sort&compress}



\end{document}